\documentclass[prl,twocolumn]{revtex4}
\usepackage{epsfig}
\usepackage{bm}

\begin{document}

\title{Hamiltonian distributed chaos in long-term solar activity}

\author{A. Bershadskii}

\affiliation{
ICAR, P.O. Box 31155, Jerusalem 91000, Israel
}

\begin{abstract}
It is shown that the long-term solar activity (represented by the dynamically covered monthly time series of the sunspot number for period 1750-2005yy) exhibits spectral properties of the Hamiltonian distributed chaos with spontaneously broken {\it time} translational symmetry. 
\end{abstract}

\maketitle

\section{Introduction}

  A magneto-hydrodynamic dynamo mechanism generates magnetic flux tubes in convective zone of the sun. Then turbulent convection captures the magnetic flux tubes and pulls some of them trough the surface of the convection zone into photosphere to form visible sunspots. A complex nonlinear interplay of the thermal convection, global rotation and magneto-hydrodynamic processes determines the dynamics of the sunspots appearance on the convection zone surface (see, for instance, Ref. \cite{b1} and references therein). It is rather difficult to infer information about the underlying physical mechanisms from the time series describing time evolution of the sunspot number on the sun surface. Therefore, recent attempts \cite{lamg} to reconstruct a low-dimensional phase portrait from the sunspot number dynamics (scalar time series), using new methods of nonlinear dynamical systems \cite{lg}, are so valuable (see also Refs. \cite{pt},\cite{lpn} and references therein). This reconstruction allows to obtain an answer for question: What type of chaos are we observing there?  The very high values of Rayleigh, Rossby and magnetic Reynolds numbers, involved in the above described processes, suggest that the chaos can be related to the Hamiltonian dynamics. Therefore, let us start from the description of the Hamiltonian distributed chaos.

\section{Hamiltonian distributed chaos}

  Chaotic behaviour of dynamical systems is often characterized by the exponential power frequency spectra \cite{oh}-\cite{fm}
$$
E(f) \propto \exp -(f/f_c)      \eqno{(1)}
$$
with $f_c = const$. 

   Nonlinear (Duffing) oscillator provides a simple example (this system is widely used in the modelling of solar activity - see, for instance, the Refs. \cite{b1},\cite{lpn},\cite{lp}-\cite{kk} and references therein): 
$$
\ddot{x} + b\dot{x} +\beta x + \alpha x^3 = A \sin  \omega t  \eqno{(2)}
$$  
At $b=0$ (negligible dissipation) it is a Hamiltonian system. At certain values of the parameters this oscillator exhibits chaotic behaviour.
Figure 1 shows (in the semi-logarithmic scales) power spectrum for the Duffing oscillator with double-well potential ($b= 0.05$, $\alpha =1$, $\beta = -1$, $\omega = 0.7$ and $A= 0.7$ \cite{spot}). The solid straight line indicates the exponential spectrum Eq. (1).  

\begin{figure} \vspace{-0.5cm}\centering
\epsfig{width=.45\textwidth,file=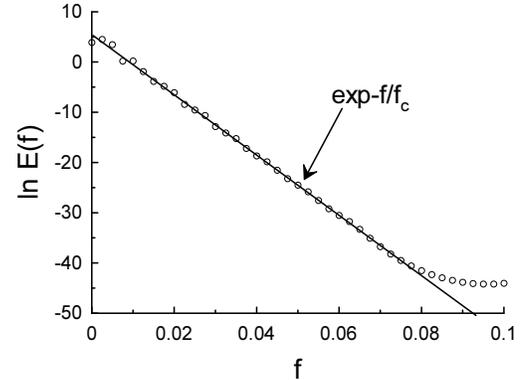} \vspace{-4cm}
\caption{Power spectrum for nonlinear (Duffing) oscillator. The solid straight line is drawn to indicate the exponential spectrum Eq. (1). }
\end{figure}

  For a more complex case of distributed chaos a weighted superposition:
$$
E(f) \propto \int_0^{\infty} P(f_c)~ \exp -(f/f_c)~ df_c   \eqno{(3)}
$$
has been used, with $P(f_c )$ as a probability distribution of $f_c$.\\

  For Hamiltonian systems the Noether's theorem relates the energy conservation to the time translational invariance (symmetry) \cite{ll},\cite{she}. Action $I$ is an adiabatic invariant in the case of spontaneously broken time translational symmetry \cite{suz} (see also next section and Ref. \cite{pt}). The dimensional considerations provide a relation between characteristic velocity ($v_c$) and frequency ($f_c$):
$$    
v_c \propto I^{1/2} f_c^{1/2}  \eqno{(4)}
$$
Normal distribution of $v_c$ results in $\chi^{2}$ distribution of $f_c$:
$$
P(f_c) \propto f_c^{-1/2} \exp-(f_c/4f_0)  \eqno{(5)}
$$
where $f_0 = const$. 

   Substitution of the $\chi^{2}$ distribution - Eq. (5) into Eq. (3) gives:
$$
E(f) \propto \exp-(f/f_0)^{1/2}  \eqno{(6)}
$$

\begin{figure} \vspace{-1.5cm}\centering
\epsfig{width=.45\textwidth,file=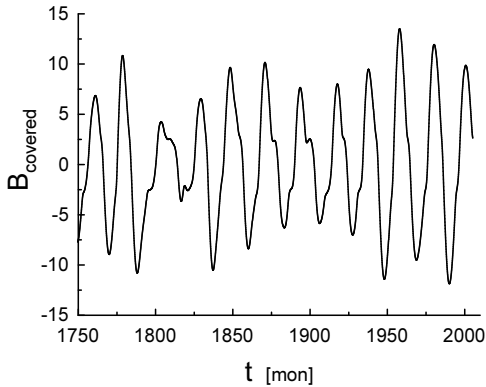} \vspace{-4cm}
\caption{Monthly time series for a surrogate of magnetic field (the data can be found at Ref. \cite{data}).}
\end{figure}
\begin{figure} \vspace{-0.5cm}\centering
\epsfig{width=.45\textwidth,file=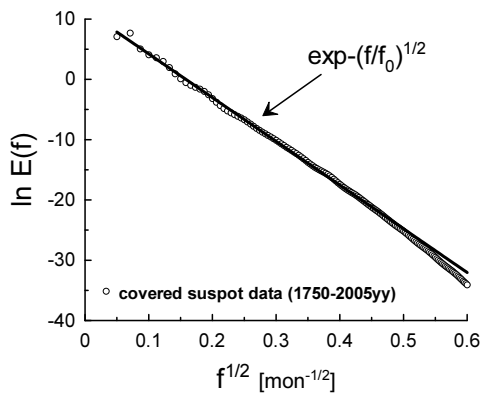} \vspace{-4cm}
\caption{Power spectrum the surrogate magnetic field fluctuations. The solid straight line indicates the stretched exponential decay Eq. (6).}
\end{figure}

\section{Long-term solar activity}

  It is shown in the recent Ref. \cite{pt} that time dynamics of a surrogate of action: the energy of the magnetic field (associated with the sunspots) divided by local (running) frequency of the solar cycle, is consistent with an adiabatic invariance of the action. \\
   
   The time series of the sunspot number is a scalar one. Therefore, in order to investigate the {\it underlying} dynamics one needs in a reconstruction of a multi-dimensional phase space. It is estimated in the Ref. \cite{lamg} that an embedding dimension equal to three is sufficient for a quantitative reconstruction of the underlying dynamics. It is well known that the solar magnetic field cycle has a polarity inversion with a period approximately equal to 11 years (that corresponds to the magnetic field cycle approximately equal to the 22 years period). Any relevant underlying dynamics should  exhibit these features, i.e. should have a corresponding symmetry group. It should be noted that the original time series of the sunspot number does not have such symmetry. The problem is to obtain a {\it cover} system (with the symmetry group) locally dynamically equivalent to the {\it image} system without the symmetry group \cite{lg}. The reconstructed phase portrait should be invariant under the inversion symmetry. This task was performed for the monthly time series of the sunspot number (for the period 1750-2005yy) in the Ref. \cite{lamg} and a corresponding time series for a surrogate of magnetic field ($\text{B}_{\text{covered}}$) is shown in the figure 2 (the data can be found at the site \cite{data}). Figure 3 shows corresponding power spectrum and the straight line is drawn to indicate correspondence to the Eq. (6) (the distributed chaos, $T_0=1/f_0=438$y). 
   
    Figure 4 shows, for comparison, power spectrum of the temperature fluctuations measured \cite{as} at the centre of a thermal convection cell at very large Rayleigh number: $Ra \simeq 3 \cdot 10^{14}$ (about Hamiltonian approach for the fluid dynamics see, for instance, the Refs \cite{she},\cite{eyi}).

\begin{figure} \vspace{-0.85cm}\centering
\epsfig{width=.45\textwidth,file=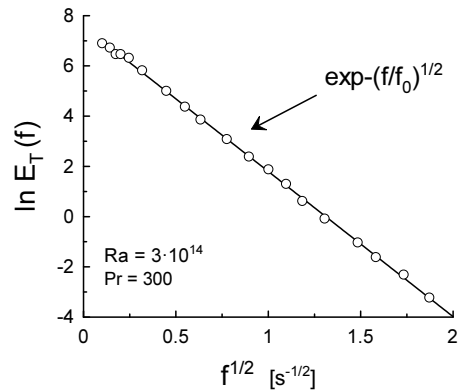} \vspace{-4.7cm}
\caption{Power spectrum of temperature fluctuations measured \cite{as} at the center of a thermal convection cell.}
\end{figure}

\section{Acknowledgement}

I thank C. Letellier, L.A. Aguirre, J. Maquet, R. Gilmore and J.C. Sprott for sharing their data.

\end{document}